\documentclass[
    ,final            
  ]
  {aipproc}

\layoutstyle{8x11double}

\newcommand{\Hp}{\rm{H}^{+}}

\newcommand{\mHt}{\rm{H}_{2}}

\newcommand{\mC}{\rm{C}}
\newcommand{\Cp}{\rm{C^{+}}}

\newcommand{\sip}{\rm{Si^{+}}}
\def\simless{\mathbin{\lower 3pt\hbox
   {$\rlap{\raise 5pt\hbox{$\char'074$}}\mathchar"7218$}}}   
\def\simgreat{\mathbin{\lower 3pt\hbox  
   {$\rlap{\raise 5pt\hbox{$\char'076$}}\mathchar"7218$}}} 
\newcommand{\hii}{\hbox{H\,{\sc ii}}\,}
\begin{document}

\title{The Influence of Metallicity on Star Formation in Protogalaxies}


\author{A.-K. Jappsen}{
  address={Canadian Institute for Theoretical Astrophysics, University of Toronto, Toronto, ON M5S 3H8, Canada},
  altaddress={Astrophysikalisches Institut Potsdam, An der Sternwarte 16, 14482 Potsdam, Germany}
 }

\author{S.~C.~O. Glover}{
  address={Astrophysikalisches Institut Potsdam, An der Sternwarte 16, 14482 Potsdam, Germany},
  altaddress={Department of Astrophysics, American Museum of Natural History, New York, NY 10024-5192, USA},
}

\author{R.~S. Klessen}{
  address={Zentrum f\"ur Astronomie der Universit\"at Heidelberg, 
Institut f\"ur Theoretische Astrophysik, Albert-Ueberle-Str.\,2, 69120 Heidelberg, Germany},
 altaddress={Astrophysikalisches Institut Potsdam, An der Sternwarte 16, 14482 Potsdam, Germany},
}

\author{M.-M. Mac Low}{
  address={Department of Astrophysics, American Museum of Natural History, New York, NY 10024-5192, USA},
  altaddress={Zentrum f\"ur Astronomie der Universit\"at Heidelberg, 
Institut f\"ur Theoretische Astrophysik, Albert-Ueberle-Str.\,2, 69120 Heidelberg, Germany},
}

\begin{abstract}
In cold dark matter cosmological models, the first stars to form are believed to do so within small protogalaxies. We wish to understand how the evolution of these early protogalaxies changes once the gas forming them has been enriched with small quantities of heavy elements, which are produced and dispersed into the intergalactic medium by the first supernovae. Our initial conditions represent protogalaxies forming within a fossil $\hii$ region, a previously ionized region that has not yet had time to cool and recombine. We study the influence of low levels of metal enrichment on the cooling and collapse of ionized gas in small protogalactic halos using three-dimensional, smoothed particle hydrodynamics (SPH) simulations that incorporate the effects of the appropriate chemical and thermal processes. Our previous simulations demonstrated that for metallicities $Z < 10^{-3} \: {Z_{\odot}}$, metal line cooling alters the density and temperature evolution of the gas by less than 1\% compared to the metal-free case at densities below $1\,\mathrm{cm^{-3}}$ and temperatures above $2000\,\mathrm{K}$. Here, we present the results of high-resolution simulations using particle splitting to improve resolution in regions of interest. These simulations allow us to address the question of whether there is a critical metallicity above which fine structure cooling from metals allows efficient fragmentation to occur, producing an initial mass function (IMF) resembling the local Salpeter IMF,  rather than only high-mass stars.
\end{abstract}

\maketitle


\section{Introduction}
Population {\sc III} stars are the first potential producers of UV photons that can
contribute to the reionization process and are the first producers of the
metals required for the formation of population {\sc II} stars. 
Metals produced by the first stars will also be injected into some fraction of the ionized volume,
and so will be present in gas falling into new or existing protogalactic halos.
The question then arises as to how this low level of metal enrichment affects the ability of the 
gas to cool and collapse. 
 
In \cite{BRO01}, the collapse of cold, metal-enriched gas in a top-hat potential was simulated.
These simulations included the effects of atomic fine structure cooling, but did not
include cooling from ${\rm H_{2}}$. In the absence of molecular cooling, 
fragmentation suggestive of a modern IMF was found to occur only for 
metallicities above a threshold value of $Z \simeq 10^{-3.5} Z_\odot$.  However,
the authors noted that  the neglect of molecular cooling could be significant.  
In \cite{OMU05}, Omukai et~al.\  argued, based on the results of their detailed one-zone 
models, that molecular cooling would indeed dominate the cooling over many orders of 
magnitude in density. Also, by starting with cold gas, the authors of \cite{BRO01}
implicitly assume that no extra entropy or energy has been added to the gas during its 
enrichment, although this is unlikely to be the case \cite{OH03}.

It is therefore important to consider whether a metallicity threshold
appears in simulations with different initial conditions.  In this
paper, we examine a different, still idealized, set of initial
conditions, using a more detailed treatment of the cooling and chemistry of the gas, 
and find no threshold. We argue that the transition from
the primordial to the modern IMF therefore depends on the conditions under
which stars form as much as on the metal abundances present.

\section{Numerical Approach}
To help us to assess the influence of metals on the fragmentation properties of
gas in small protogalactic halos, we have performed a number of numerical 
simulations. Here we present a portion of our results. Detailed discussion of
the full set of simulations can be found in \cite{jmgk07} (hereafter paper V).
Our study requires us to follow the collapse of the gas over many orders of
magnitude in density, and therefore the problem is well-suited for the
use of a Lagrangian numerical method.  We use 
 SPH, a Lagrangian method described in \cite{BEN90}, \cite{MON92} 
and \cite{MON05}, in which the fluid is followed with an ensemble of particles.  
Fluid properties at each point are computed by averaging over neighboring particles. 
The calculations presented in this paper use version 1 of the parallel SPH
code GADGET \cite{SPR01}.
In order to achieve a higher mass resolution, we refine the mass of the gas particles using 
a method known as particle splitting \cite{KIT02}. In \cite{jgkm07}, hereafter paper II,
we presented low-resolution simulations that show that the density increases most rapidly close to the center of the dark matter halo. On-the-fly splitting with two levels of refinement at two 
different radii provides us with the highest mass resolution at the region of interest. In the region of collapse we can resolve Jeans masses down to 1.5~$M_{\odot}$, as compared to $M_{\mathrm{res}} = 200~M_{\odot}$ in the low resolution simulations. 
We discuss the details of the implementation of the particle splitting algorithm in paper V.

As we want to represent gas that has collapsed beyond the
resolution limit of the simulation in a numerically robust manner, we
have modified the code to allow it to create sink particles --
massive, non-gaseous particles, designed to represent dense cores,
that can accrete gas from their surroundings but otherwise interact
only via gravity \citep{BAT95}. 
Sink particles are created once the density rises above $10^4\,\mathrm{cm^{-3}}$ 
and are endowed with an accretion radius of $0.3\,\mathrm{pc}$. On every time step, 
any gas within this accretion radius that is gravitationally bound to the sink particle 
is accreted by it.  The design and implementation of our sink particle algorithm is 
discussed in more detail in \cite{JAP05}. 

\section{Chemistry and Cooling}
We have further modified GADGET to allow us to follow the non-equilibrium
chemistry of the major coolants in both primordial and low-metallicity gas.
Our chemical model is presented in \cite{gj07} and \cite{glo07} (hereafter
papers I and III). Provided that carbon and oxygen are amongst 
the most abundant metals, the major coolants will be largely the same as in 
local atomic and molecular gas, namely $\mHt$, HD, C, $\Cp$, O, Si, $\sip$, CO, 
OH and ${\rm H_{2}O}$ \citep{OMU05}. We therefore follow the abundances of 
these ten species, together with an additional 29 species that play important 
roles in determining the abundances of one or more of these coolants. 
Our chemical network  contains a total of 189 collisional 
gas-phase reactions between these 39 species, as well as 7 grain surface 
reactions, 43 reactions involving the photoionization or photodissociation of 
chemical species by ultraviolet radiation, and 8 reactions involving cosmic 
rays. For simplicity, in the simulations presented in this paper we do not include
the effects of dust, UV radiation or cosmic rays, and so use a simplified 
version of the model that contains only the collisional reactions. 
The thermal evolution of the gas in our simulations is modelled using a
cooling function that includes the effects of atomic fine structure cooling
from $\mC$, $\Cp$, O, Si and $\sip$, rotational and vibrational cooling
from $\mHt$, HD, CO and ${\rm H_{2}O}$, Lyman-$\alpha$ cooling,
Compton cooling, and $\Hp$ recombination cooling, as well as a number
of other processes of lesser importance.
For additional details of our chemical networks and cooling function, please consult
papers I and III; further details of their implementation within Gadget are also given
in paper II.

\section{Initial Conditions}
Our initial conditions are based on those used in paper II,
although collapse is followed to much higher density in the current work.  
We study protogalaxies forming from fully ionized gas with initial
temperature $T_g = 10^4$~K, representing a fossil H~{\sc ii} region
\citep{OH03} polluted by supernovae from the ionizing object. 
We model one such halo by using a fixed background potential with a spherically symmetric density profile \cite{NAV97}. We study the evolution both of zero-metallicity, primordial gas and of 
metal-enriched gas with a metallicity ${Z} = 10^{-3} {Z_{\odot}}$.  To simplify the discussion of our simulation
results, we take as a fiducial example a halo with a dark matter mass 
$M_{\rm dm} = 7.8 \times 10^{5} \: {M_{\odot}}$, an initial redshift 
$z_{i} = 25$, and a spin parameter $\lambda = 0.05$. For this example halo, the virial temperature 
$T_{\rm vir} = 1900 \, \mathrm{K}$, the virial radius $r_{\rm vir} = 0.1 \, {\rm kpc}$,
and the truncation radius $r_{{t}} = 0.49\:\mathrm{kpc}$, where both
radii are given in physical units. The scale radius $r_{{s}}$ of this 
example halo is $29\,\mathrm{pc}$, and the full computational volume is a box 
of side length $1\,\mathrm{kpc}$. The gas mass is $M_{{g}}=0.19\,M_{\mathrm{dm}}$.  

\section{Results and Discussion}

In the simulations presented here we focus on gas with a metallicity 
${\rm Z} = 10^{-3} \: {Z_{\odot}}$. This metallicity
is at the upper limit of the range of values proposed 
for the so-called critical metallicity ${Z_{\rm crit}}$,
the value of the metallicity at which efficient fragmentation 
and low-mass star formation is hypothesized to first occur 
\citep{BRO01,OMU05,BL03,SCH02}. It is also comparable to the 
globally averaged metallicity produced by the sources responsible 
for reionization \citep[see e.g.][]{RIC04}.

\begin{figure}
  \includegraphics[width=.47\textwidth]{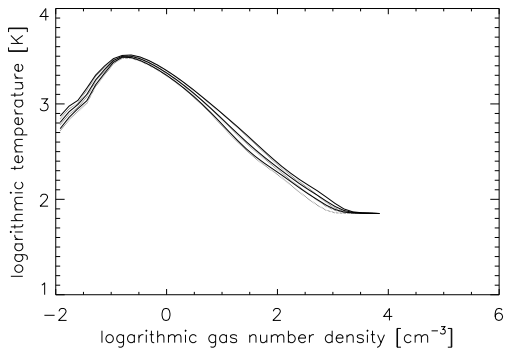}
  \caption{Gas temperature vs. number density for the runs with metallicities 0.0 ({\it solid line}), $10^{-3}$ ({\it dotted line}), and $10^{-1} \;Z_{\odot}$ ({\it dashed line}). The spin parameter of the runs is 0.05. The time of the plot is half a Hubble time. The thin lines show the 1-$\sigma$ deviation.}
  \label{fig1} 
\end{figure}

\begin{figure}
  \includegraphics[width=.23\textwidth]{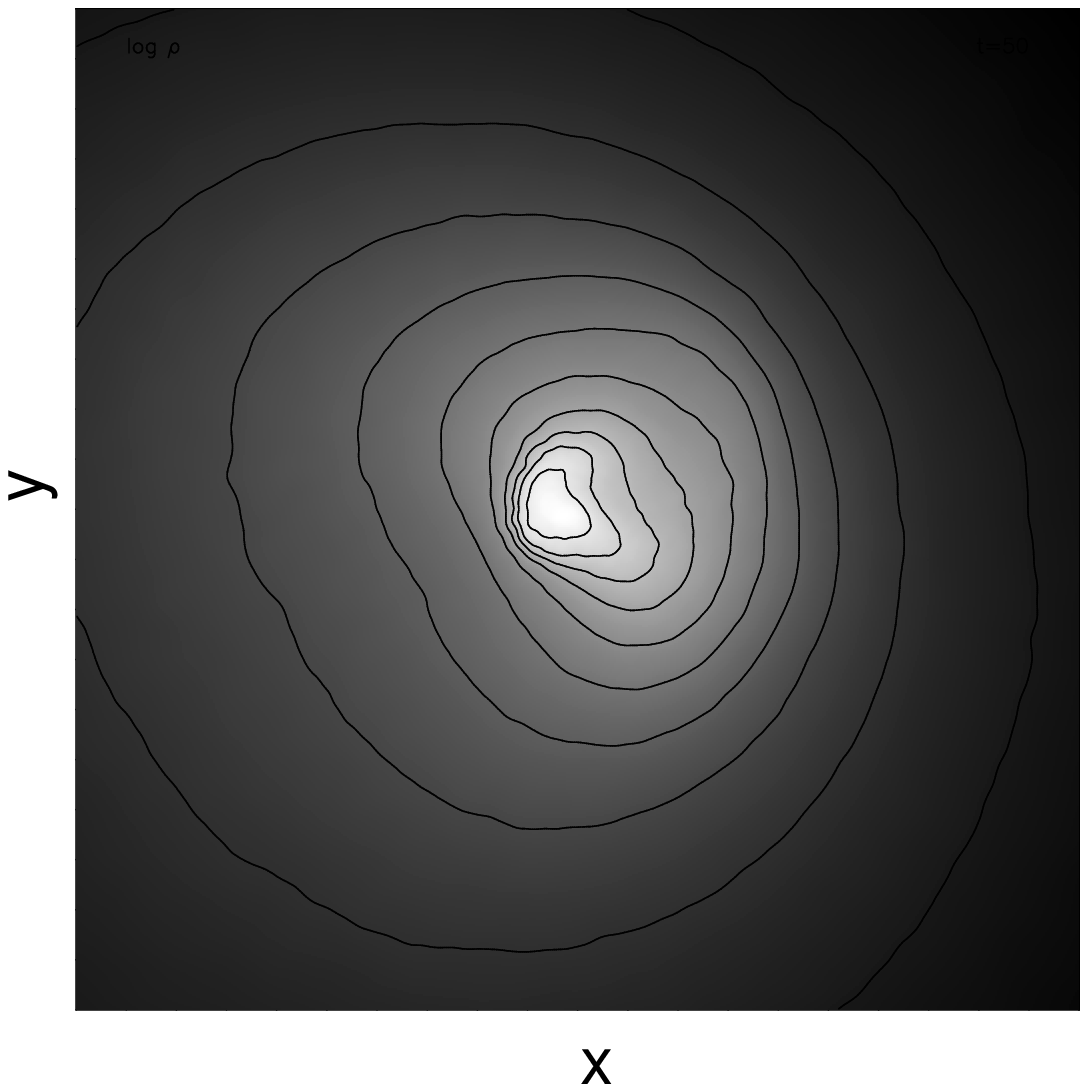}
  \includegraphics[width=.245\textwidth]{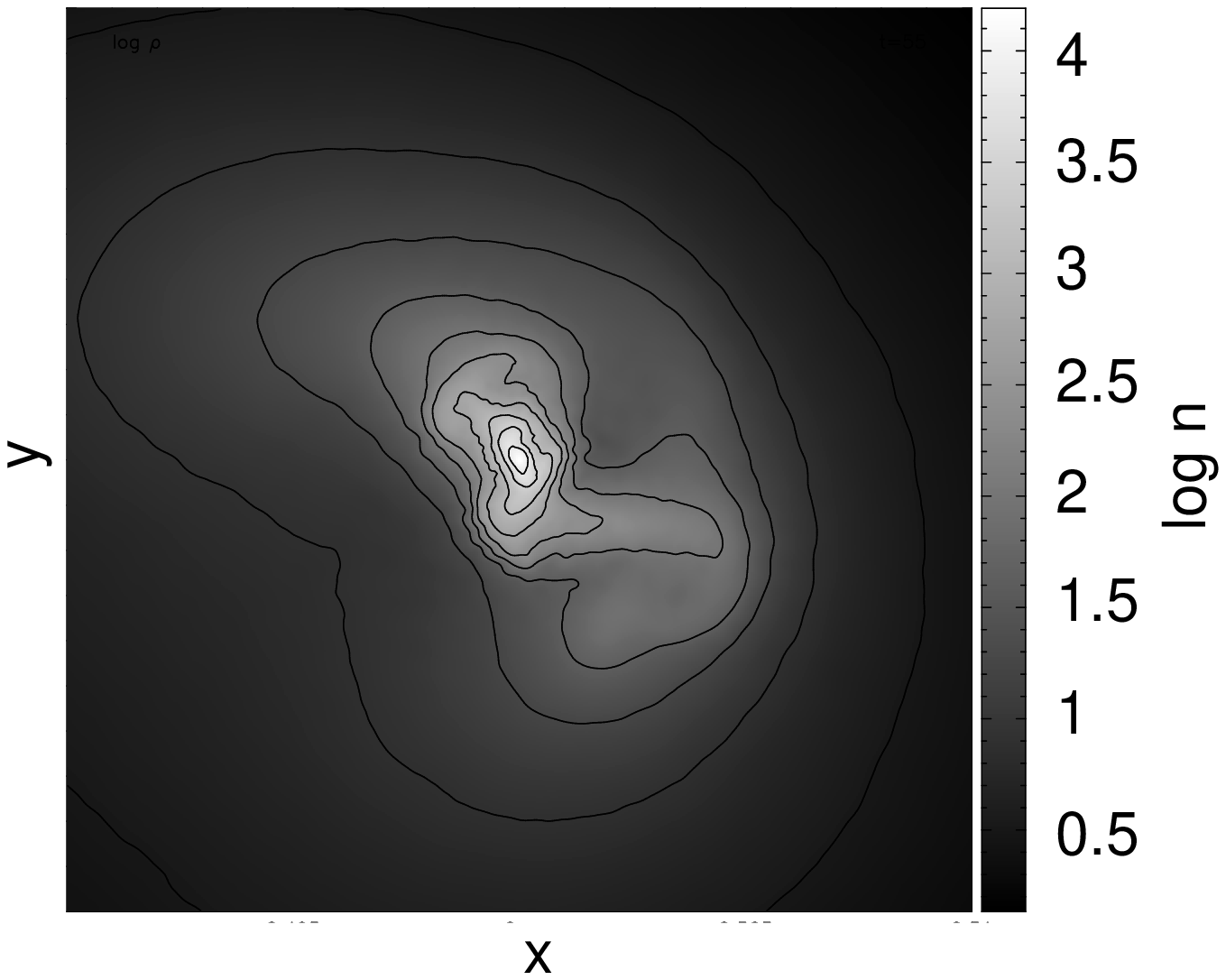}
 \caption{Cut at $z = 0.5 \,\mathrm{kpc}$ through the gas number density for the run with zero metallicity ({\it left panel}) and the run with a metallicity of $ {Z} = 10^{-3} {Z_{\odot}}$ ({\it right panel}). We show a box with a size of 20 pc.}
 \label{fig2}
\end{figure}

In Figure~\ref{fig1} we compare the temperature-density evolution of the run with primordial gas and the run with enriched gas. The two runs differ by less than 10\% up to densities of $n = 10^4 \: {\rm cm^{-3}}$. 
This result demonstrates that the cooling of the gas is barely influenced 
by the presence of metals: fine-structure cooling contributes 
only marginally to the total cooling rate. $\mHt$ is the dominant coolant,
rather than metal fine structure lines.
At first sight, the fact that fine structure cooling from metals has little impact on 
the thermal or dynamical evolution of the gas at  metallicities below
$0.1 \: {Z_{\odot}}$ is somewhat surprising, given that it was previously
found that gas with a metallicity of only ${Z} = 10^{-3} \: {Z_{\odot}}$ could cool 
rapidly and fragment even in the complete absence of molecular hydrogen
\cite{BRO01}. However, this difference in conclusions appears to be a
consequence of the different initial conditions used in the two sets of simulations,
and the difference in the physics included. The simulations presented in \cite{BRO01}
neglected molecular cooling, and started with cold ($T_{\mathrm{initial}}= 200\,\mathrm{K}$), 
low ionization ($x_{e}=10^{-4}$) gas in a flat, top-hat dark matter potential lacking a central condensation, although with local density perturbations. In these simulations, 
the gas cools, collapses and fragments only if the metallicity is high enough to 
render the cooling time shorter the the local dynamical timescale. With our arguably more 
realistic initial conditions, and, crucially, with our inclusion of ${\rm H_{2}}$, we find no 
metallicity threshold -- ${\rm H_{2}}$ cooling alone is sufficient to allow the gas
to collapse, but the collapsing gas does not fragment (see Figure~\ref{fig2} and \cite{jkgm07}). 
In \cite{jmgk07},  we present a set of additional runs with metallicities up to 
${Z} = 10^{-1} {Z_{\odot}}$ that confirm this result. We conclude that the transition 
from the primordial to the modern-day IMF therefore depends on the conditions under
which stars form as much as on the metal abundances present.  The actual conditions 
that are appropriate for second generation star formation still need to be determined in 
detail by observation and modeling of galaxy formation.


\begin{theacknowledgments}
RSK and AKJ acknowledge support from the Emmy Noether Program of the DFG (grant KL1385/1). M-MML and SCOG were supported in part by AST03-07793 and M-MML was supported in part by a DAAD stipend. The simulations described here were performed on the PC cluster "sanssouci" at the AIP and the cluster "McKenzie" at CITA.
\end{theacknowledgments}



\bibliographystyle{aipproc}   




\end{document}